\documentclass[twocolumn,amsmath,amssymb,prl,showpacs]{revtex4}
\usepackage{setspace}
\usepackage{graphicx}
\usepackage{amsmath}
\usepackage{amssymb}
\usepackage{latexsym}

\usepackage{natbib}
\begin{document}
%\singlespacing
%\onehalfspacing
%\doublespacing

\title{Exploiting soliton decay and phase fluctuations in atom chip interferometry of Bose-Einstein condensates}
\author{R.G. Scott, T.E. Judd, T.M. Fromhold}
\affiliation{School of Physics and Astronomy, University of Nottingham, Nottingham, NG7 2RD, United Kingdom.}
\date{5/10/07}
%\maketitle

%\onehalfspacing 
%\doublespacing
%\singlespacing

\begin{abstract}
We show that the decay of a soliton into vortices provides a mechanism for measuring the initial phase difference between two merging Bose-Einstein condensates. At very low temperatures, the mechanism is resonant, operating only when the clouds start in anti-phase. But at higher temperatures, phase fluctuations trigger vortex production over a wide range of initial relative phase, as observed in recent experiments at MIT. Choosing the merge time to maximize the number of vortices created makes the interferometer highly sensitive to spatially varying phase patterns and hence atomic movement.
\end{abstract}


\maketitle

Recent experiments~\cite{kettnew} have successfully performed interferometry of two Bose-Einstein condensates (BECs) in an atom chip trap. The two BECs were merged over a precisely controlled timescale and then allowed to expand freely. The size of the cloud after expansion was found to increase with the initial phase difference, $\Delta$, between the two BECs, being maximal when $\Delta = \pi$. Consequently, measurements of the cloud size following expansion can be used to determine $\Delta$, and hence form the basis of an atom interferometer. However, the physical mechanism that makes the expanded cloud size depend on $\Delta$ is unknown. As a result, many interesting questions remain unanswered, for example concerning the variation of expanded cloud size with merging time~\cite{kettnew}. In addition, for the temperatures and extremely elongated geometry of the BECs studied in the experiments, phase fluctuations are predicted~\cite{kettnew2,petrov}, but it is unclear how this affects the operation of the interferometer. 

%Potentially, this phase measurement technique could be combined with \textit{in situ} imaging~\cite{kettinsit}, and hence a single BEC could be used for multiple operations of the interferometer. The trapping potentials and detection optics could be integrated into a single atom chip~\cite{steinmetz}, creating a compact and portable detector of force or motion. 

%Recent experiments~\cite{kettnew} have successfully performed interferometry of Bose-Einstein condensates (BECs) on an atom chip. By merging two elongated BECs over a precisely controlled timescale, their initial relative phase was inferred by measuring the size of the final combined cloud. It was found that a relative phase of $\pi$ led to greater expansion of the merged BEC than zero relative phase. By combining this phase measurement technique with \textit{in situ} imaging~\cite{kettinsit}, it may be possible to use a single BEC for multiple operations of the interferomter. Furthermore, the trapping potentials and detection optics could be integrated into a single atom chip~\cite{steinmetz}, creating a compact and portable detector of force or motion. However, the physical reason why expansion rate depends on phase shift has not hitherto been understood. As a result, many open questions remain unanswered, in particular about role of the merging time~\cite{kettnew}. Furthermore, for the temperatures and extremely elongated geometry of the experiment, phase fluctuations are predicted along the long axis of the BEC~\cite{kettnew2,petrov}. It is unclear how this effects the function of the interferometer. 

In this Letter, we show that vortex production is the physical mechanism that causes the size of the atom cloud after expansion to vary with $\Delta$. When $\Delta = \pi$, a soliton forms between the two merging BECs and then decays into vortices. The additional momentum associated with the vortices drives an enhanced spreading of the cloud when it is released from the chip trap. We investigate the role of the merging time on the creation of dynamical excitations and predict that the number of vortices produced will be maximal for a merge time of $\sim5$ ms. Surprisingly, the finite temperature phase fluctuations~\cite{kettnew2,petrov} characteristic of elongated BECs can actually enhance the sensitivity of the interferometer. This is because, in the absence of phase fluctuations, vortex production is a sharply resonant process that only occurs when $\Delta = \pi$. However, the presence of phase fluctuations causes the number of vortices generated by the merging to change more smoothly with $\Delta$. Consequently, the action of the interferometer can be controlled by varying the temperature and/or geometry of the BEC.

In the experiments~\cite{kettnew}, a $^{23}$Na BEC containing $N = 4 \times 10^{5}$ atoms is held in an atom chip trap produced by three current-carrying wires and an external bias field. As expected for atom chips, the trap frequencies are very high perpendicular to the trapping wire ($\omega_x = \omega_y = 2 \pi \times 1000$ rad s$^{-1}$), but low along the wire ($\omega_z = 2 \pi \times 9$ rad s$^{-1}$), thus creating an exceedingly elongated cigar-shaped BEC of peak density $n = 2.8\times10^{14}$ cm$^{-3}$. For aspect ratios this extreme, finite temperature phase fluctuations prevent global coherence of the atom cloud, even at temperatures $T << T_{C}$, the transition temperature~\cite{petrov}. In the limit of a truly one-dimensional gas, phase coherence is completely lost, even at $T=0$~\cite{Hofferberth}. In the MIT experiments, $T \approx 0.5$ $T_{C}$, which is far higher than the threshold of $0.1$ $T_{C}$ above which phase fluctuations are expected~\cite{kettnew}. 

\begin{figure}[tbp]
\centering
\includegraphics[keepaspectratio=true, scale=0.47]{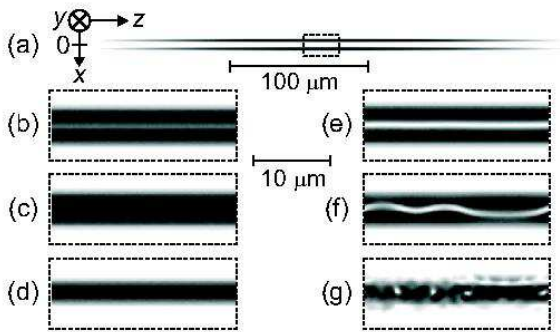}
\caption{(a) Atom density profile of the two BECs in the $y=0$ plane (axes inset) at $t=0$. (b)-(g) Density profiles within the region enclosed by the dashed rectangle in (a) at key stages of the merging process ($\tau = 5$ ms) calculated for $\Delta = 0$ at $t=3$ ms (b), 4 ms (c), 5 ms (d), and for $\Delta=\pi$ at $t=3$ ms (e), 4 ms (f), and 5 ms (g). Upper [lower] horizontal bars show scales in (a) [(b)-(g)].}
\label{f1}
\end{figure}

By applying an oscillating rf field from a fourth wire on the atom chip~\cite{kruger,kettold}, the single-well potential is deformed smoothly into a double-well potential, thus splitting the BEC along its short ($x$) axis [Fig.~\ref{f1}(a)]. We assume that this process is performed adiabatically, and so begin our simulations with two equilibrium groundstate BECs, each containing $2 \times 10^{5}$ atoms, calculated by solving the three-dimensional Gross-Pitaevskii equation in imaginary time~\cite{chiofalo,meotago}. In the experiments~\cite{kettnew}, $\Delta$ is set by holding the atoms for a pre-determined time before beginning the merging process. Due to small asymmetries in the trapping potential, there is a reproducible difference between the chemical potentials of the two clouds, which causes a non-vanishing and measurable phase evolution rate~\cite{kettold}. In our calculations, at the start of the merging process (time $t=0$) we set the phase of the BEC to $0$ for $x<0$ and $\Delta$ for $x>0$. The BECs are then merged by smoothly transforming the double-well potential back into a single-well potential over a known merging time $\tau$, as in experiment~\cite{kettnew}.
%Firstly, we take $\Delta$ to be spatially invariant, but later consider phase profiles with some $z-$dependence. 

Throughout the merging process, we model the three-dimensional wavefunction by an expansion over plane-wave basis states~\cite{meotago}, with a maximum cutoff wavevector, $k_{m}$, chosen to prevent Fourier aliasing~\cite{norrie}. In the absence of fluctuations, this method is equivalent to solving the projected Gross-Pitaevskii equation~\cite{davis}. The Gross-Pitaevskii equation has been very successful in calculating equilibrium states and modeling mean-field dynamics, but it fails to describe spontaneous processes, such as the formation of a scattering halo and the associated depletion of the BEC. We include such effects in our calculations by adding random complex noise to each plane wave mode~\cite{meotago,norrieprl}. The amplitude of these quantum fluctuations has a Gaussian distribution, with an average value of half a particle. This approach is known as the truncated Wigner method~\cite{meotago,norrieprl,steel,sinatra}. 
%However, the truncated Wigner method is still a zero temperature theory. 

In order to model the finite temperature phase fluctuations characteristic of elongated BECs, we add extra thermal fluctuations to low energy modes of the BEC below a small wavevector cutoff, $k_{s} = 1.3 \times 10^6$ m$^{-1}$, which is just sufficient to describe the overall curvature of the initial BEC wavefunction. To allow this extra noise to reach dynamical equilibrium, we propagate the noisy wavefunction in time, using a plane-wave basis with wavevectors up to $k_{s}$. Hence, we create macroscopic thermal fluctuations in the low energy modes, with an approximate Bose-Einstein distribution~\cite{davis,poles}. Finally, when the mode populations have stabilized, we increase the cutoff wavevector to $k_{m}$, add quantum fluctuations to all modes, and begin the merging process~\cite{foot1}. This is a finite temperature classical field method~\cite{davis,poles,blakie}. Here, we present simulations without fluctuations, including quantum fluctuations only, and including both quantum and thermal fluctuations. We compare our results for each method to isolate the effect of the mean field, the quantum noise, and the thermal noise, on the performance of the interferometer. 

%In order to model the finite temperature phase fluctuations characteristic of elongated BECs, we add extra thermal fluctuations to low energy modes of the BEC below a small wavevector cutoff, $k_{s} = 1.3 \times 10^6$ m$^{-1}$, which is just sufficient to describe the overall curvature of the initial BEC wavefunction. To allow this extra noise to reach dynamical equlibrium, we propagate the noisy wavefunction in time, using a plane-wave basis with wavevectors up to $k_{s}$. Hence, we create macroscopic thermal fluctuations in the low energy modes, with an approximate Bose-Einstein distribution~\cite{davis,poles}. Finally, when the mode populations have stabilized, we increase the cutoff wavevector to $k_{m}$, add quantum fluctuations to all modes, and begin the merging process~\footnoteremember{myfootnote}{Strictly, we should keep our cutoff wavevector small and use a harmonic oscillator basis to ensure macroscopic thermal occupation of the BEC modes~\cite{davis}. Consequently, we do not attempt to ascribe a specific temperature value to our simulations.}. This is a finite temperature classical field method~\cite{davis,poles,blakie}. Here, we present simulations without fluctuations, including quantum fluctuations only, and including both quantum and thermal fluctuations. We compare our results for each method to isolate the effect of the mean field, the quantum noise, and the thermal noise, on the performance of the interferometer. 

We begin by presenting simulations without fluctuations, taking $\tau = 5$ ms (Fig.~\ref{f1}). For clarity, Figs.~\ref{f1}(b-g) only show the central part of the atom cloud, within the region enclosed by the dashed rectangle in Fig.~\ref{f1}(a), at various times during the merging process. When $\Delta = 0$, the two clouds combine gradually [Figs.~\ref{f1}(b,c)] to form a final BEC [Fig.~\ref{f1}(d)] with a smooth density profile, which is very similar to the ground state of the single-well potential. However, the dynamics are much more complicated if $\Delta = \pi$. In this case, the phase shift creates a density node between the two approaching BECs, which evolves into a soliton~\cite{bongs}, shown by the central white band in Fig.~\ref{f1}(e). Since this excitation is unstable, it bends [Fig.~\ref{f1}(f)] and then decays into vortices [Fig.~\ref{f1}(g)], via the snake instability~\cite{anderson,dutton}. 
%Due to the addition of multiple topological excitations, the final cloud has a disrupted and fragmented appearance.
 
The addition of a soliton increases the kinetic energy of the wavefunction by $\sim\hbar \omega_x = 7\times10^{-31}$ J per particle. When the soliton decays, this energy is transformed into the kinetic energy of the vortices. Since the BEC is in the Thomas-Fermi regime, in the absence of solitons or vortices the energy available to expand the cloud is simply the mean-field energy, which is approximately $0.5n \times 4\pi\hbar^{2}a/m = 15\times10^{-31}$ J per particle, where $a=2.9$ nm is the s-wave scattering length and $m$ is the mass of a $^{23}$Na atom. The soliton increases this energy by $\sim50\%$, causing an enhanced spreading of the cloud when the BEC is released from the trap, as seen in experiment~\cite{kettnew}. 
%Consequently, the soliton and resulting vortices increase the mean absolute wavevector, and hence the expanded cloud width, by a factor of $\sqrt{1.5}\approx1.2$, as observed in experiment~\cite{kettnew}.

\begin{figure}[tbp]
\centering
\includegraphics[keepaspectratio=true, scale=0.45]{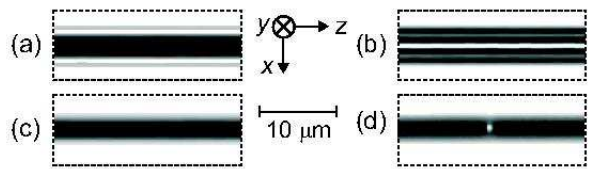}
\caption{Atom density profiles within the region of the $y=0$ plane (axes inset) shown by the dashed rectangle in Fig. 1(a), at $t = \tau=1$ ms for $\Delta = 0$ (a) and $\pi$ (b), and at $t = \tau = 50$ ms for $\Delta = 0$ (c) and $\pi$ (d). Horizontal bar shows scale.}
\label{f2}
\end{figure}

In the experiments~\cite{kettnew}, changing $\Delta$ was found to have little effect on the expanded cloud width for either very small ($\lesssim 5$ ms) or very large ($\gtrsim 50$ ms) values of $\tau$. Our simulations reveal similar behavior and enable us to identify the underlying physical reasons, which are different in the low and high $\tau$ regimes. When $\tau$ is very small, the merging process is dominated by the impact of the collision between the BECs, irrespective of the value of $\Delta$. To illustrate this, Fig.~\ref{f2} shows the merged atom density profiles for $t = \tau = 1$ ms and $\Delta = 0$ (a) and $\pi$ (b), obtained from simulations without fluctuations. Both density profiles have modulations in the $x$-direction, caused by interference between the colliding BECs. In both cases, the subsequent expansion of the merged BEC is largely determined by the momentum of the colliding atoms, which masks any enhanced spreading of the cloud resulting from the soliton in Fig.~\ref{f2}(b). When $\tau$ is very large, the BECs recombine at low speed. As a result, the merged density profile for $\Delta = 0$, shown in Fig.~\ref{f2}(c) for $t = \tau =50$ ms, is similar to the ground state of the single-well potential. In this large $\tau$ regime, the soliton formed when $\Delta=\pi$ decays before the BECs are fully merged, and consequently few vortices are produced. For example, the density profile for $t=\tau=50$ ms and $\Delta = \pi$ [Fig.~\ref{f2}(d)] is almost identical to the equivalent image for $\Delta=0$ [Fig.~\ref{f2}(c)], except that it contains a single vortex. Since this vortex has a negligible effect on the subsequent expansion of the atom cloud, in experiment the width of the expanded cloud is independent of $\Delta$. 
%The dynamics depend most strongly on $\Delta$ for intermediate $\tau \approx 5$ ms, which generates smooth density profiles for $\Delta = 0$ [Fig.~\ref{f1}(d)], but large numbers of vortices for $\Delta = \pi$ [Fig.~\ref{f1}(g)].

\begin{figure}[tbp]
\centering
\includegraphics[keepaspectratio=true, scale=0.46]{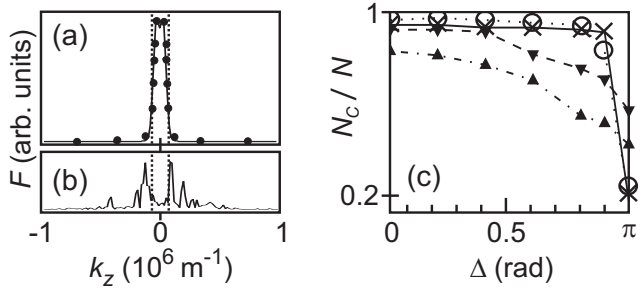}
%\vspace{-3.5cm}
\caption{Fourier power spectra, $F\left(k_z\right)$, obtained from simulations without fluctuations at $t = \tau = 5$ ms, for $\Delta = 0$ [solid curve in (a)], $0.8 \pi$ [filled circles in (a)], and $\pi$ [solid curve in (b)]. The vertical dotted lines indicate $k_z = \pm k_c$. (c) $N_c/N$ versus $\Delta$ curves ($t = \tau = 5$ ms) obtained from simulations without fluctuations (solid curve with crosses), including quantum fluctuations (dotted curve with circles), and including both quantum and thermal fluctuations at $T=T_1$ (dashed curve with downward pointing triangles) and $T_2$ (dot-dashed curve with upward pointing triangles).}
\label{f3}
\end{figure}

Since the many vortices formed for intermediate $\tau$ ($\sim 5$ ms) are described by high momentum plane wave modes, we can quantify their production by taking Fourier transforms of the merged BEC wavefunction. For simplicity, we integrate the atom density along both the $x-$ and $y-$directions and then calculate the one-dimensional Fourier power spectrum, $F\left(k_z\right)$~\cite{foot2}. For $t = \tau = 5$ ms and $\Delta = 0$, no vortices are present in the BEC [Fig.~\ref{f1}(d)] and, consequently, $F\left(k_z\right)$ [solid curve in Fig.~\ref{f3}(a)] is sharply peaked around $k_z=0$. In contrast, for $\Delta = \pi$, $F\left(k_z\right)$ is far broader [solid curve in Fig.~\ref{f3}(b)], because the BEC contains many vortices [Fig.~\ref{f1}(g)]. 

%Since the many vortices formed for intermediate $\tau$ ($\sim 5$ ms) are described by high momentum plane wave modes, we can quantify their production by taking Fourier transforms of the merged BEC wavefunction. For simplicity, we integrate the atom density along both the $x-$ and $y-$directions to calculate the one-dimensional Fourier power spectrum, $F\left(k_z\right)$~\footnote{Since the vortices are approximately cylindrically symmetrical along the $y-$direction, their associated Fourier components are almost identical in the $x-$ and $z-$directions.}. For $t = \tau = 5$ ms and $\Delta = 0$, no vortices are present in the BEC [Fig.~\ref{f1}(d)] and, consequently, $F\left(k_z\right)$ [solid curve in Fig.~\ref{f3}(a)] is sharply peaked around $k_z=0$. In contrast, for $\Delta = \pi$, $F\left(k_z\right)$ is far broader [solid curve in Fig.~\ref{f3}(b)], because the BEC contains many vortices [Fig.~\ref{f1}(g)]. 
%When the BEC is released from the trapping potential, the atoms in high momentum modes, associated with the vortices, travel quickly away from the trap center causing enhanced spreading of the cloud.

Our simulations without fluctuations predict sharply resonant vortex production when $\Delta = \pi$. This can be seen from Fig.~\ref{f3}(a), which shows that $F\left(k_z\right)$ for $\Delta = 0.8\pi$ (filled circles) is almost identical to the corresponding curve for $\Delta = 0$ (solid curve). To quantify this resonant behavior, we consider the number of atoms, $N_c = \int^{k_c}_{-k_c}F\left(k_z\right) dk_z$, with small wavevectors of magnitude $\left|k_z\right| \leq k_c=7 \times 10^4$ m$^{-1}$ in the merged BEC [the dotted lines in Figs.~\ref{f3}(a) and (b) indicate $k_z = \pm k_c$]. The quantity $N_c$ is close to $N$ for vortex-free BECs with smooth density profiles, whose $F\left(k_z\right)$ curve is sharply peaked within $-k_c < k_z < k_c$ [Fig.~\ref{f3}(a)]. However, $N_c$ drops rapidly as topological excitations appear and broaden $F\left(k_z\right)$ [Fig.~\ref{f3}(b)]. In Fig.~\ref{f3}(c), we plot $N_c/N$ as a function of $\Delta$ for merged BECs with $t=\tau=5$ ms. For the simulations without fluctuations (solid curve with crosses), $N_c/N$ decreases rapidly for $\Delta > 0.9 \pi$. When $\Delta = \pi$, $N_c/N << 1$ because almost all of the atoms have large wavevectors, and therefore travel quickly from the trap center when released from the confining potential. Consequently, the resonant production of vortices should reveal itself as a sudden increase in the width of the expanded cloud as $\Delta$ approaches $\pi$. Simulations including quantum fluctuations [dotted curve with circles in Fig.~\ref{f3}(c)] show similar resonant behaviour. However, the experiments found a far smoother variation of the expanded cloud width with $\Delta$~\cite{kettnew}. We therefore conclude that \emph{finite temperature} phase fluctuations must also be considered to explain the experiments. 

%In an attempt to reproduce the experimental findings, we repeated our simulations including quantum fluctuations. The resulting variation of $N_c$ with $\Delta$ is shown by the dotted curve with squares in Fig.~\ref{f3}(c). As for the calculations without fluctuations, $N_c$ deceases rapidly for $\Delta > 0.9 \pi$, due to the resonant production of vortices when $\Delta = \pi$. Since the inclusion of quantum fluctuations does not significantly alter $N_c\left(\Delta\right)$, we conclude that finite temperature phase fluctuations must also be considered. 
%In the limit of zero temperature, phase fluctuations are caused by the Heisenberg uncertainty principle. The uncertainty in phase $\Delta \phi \approx 1/\Delta N$, where $\Delta N\approx \sqrt{N}$ is the uncertainty in atom number~\cite{kettold}. However, as the temperature rises towards the transition temperature $T_{C}$, phase fluctuations increase dramatically, eventually causing the BEC to break up into several quasi-condensates with random phase~\cite{Hofferberth}. In very elongated geometries, this effect may occur at temperatures far below $T_{C}$, because thermal excitations of low energy longitudinal modes lead to long wavelength phase fluctuations~\cite{petrov}. 
%In the experiments of Ref.~\cite{kettnew}, the BEC temperature was $0.5$ $T_{C}$, which far exceeds the temperature of $0.1$ $T_{C}$ for the onset of phase fluctuations. 

\begin{figure}[tbp]
\centering
\includegraphics[keepaspectratio=true, scale=0.56]{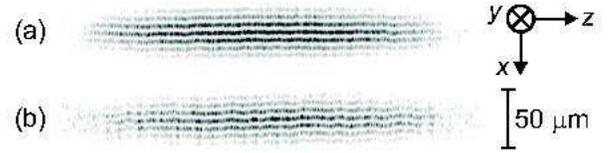}
\caption{Atom density profiles in the $y=0$ plane (axes inset) showing interference between the two BECs after 8 ms time-of-flight with $\Delta = 0$ at $T = T_1$ (a) and $T_2$ (b). Vertical bar shows scale.}
\label{f4}
\end{figure}

To elucidate the effects of finite temperature, we first study phase fluctuations by abruptly releasing two BECs from the double-well potential at $t=0$ and allowing them to expand into each other for 8 ms, as in experiment~\cite{kettnew2}. Phase fluctuations can then be observed as kinks in the resulting interference fringes. Figure~\ref{f4}(a) shows an interference pattern in the $y=0$ plane produced by two BECs which, on average, have two atoms of extra noise in each low energy mode with $k < k_{s}$. We refer to the corresponding temperature $T$ as $T_1$~\cite{foot1}. Although the fringes are very clear, their $x$-position varies along the $z-$direction due to phase fluctuations, which, in effect, make $\Delta$ weakly $z-$dependent. As we raise $T$ by adding more noise, the magnitude of the phase fluctuations increases and their mean wavelength decreases. Consequently, the interference pattern obtained for BECs with four atoms of extra noise per mode (corresponding to $T = T_2 > T_1$) has larger kinks in the fringes, which occur over shorter distances along $z$ [Fig.~\ref{f4}(b)].

We now calculate $N_c/N$ as a function of $\Delta$ for merging BECs at $T = T_1$ and $T_2$. Since each individual simulation is noisy, we repeat the merger five times for each $\Delta$ value and average the results. In contrast to our $T=0$ calculations, our simulations at $T=T_1$ reveal the production of vortices for all $\Delta \gtrsim 0.5\pi$. As a result, the corresponding $N_c/N$ versus $\Delta$ curve [dashed with downward pointing triangles in Fig.~\ref{f3}(c)] is almost constant for $\Delta \lesssim 0.5\pi$, but decreases approximately linearly with increasing $\Delta > 0.5\pi$. At the higher $T = T_2$, $N_c/N$ varies more smoothly with $\Delta$ [dot-dashed curve with upward pointing triangles in Fig.~\ref{f3}(c)], as observed in experiment~\cite{kettnew}. In the limit of very high $T$, the two BECs break up into multiple quasi-condensates, and their initial relative phase at any particular $z$ co-ordinate is essentially random. In these circumstances, the number of vortices produced is independent of $\Delta$, and so the interferometer no longer functions.

\begin{figure}[tbp]
\centering
\includegraphics[keepaspectratio=true, scale=0.57]{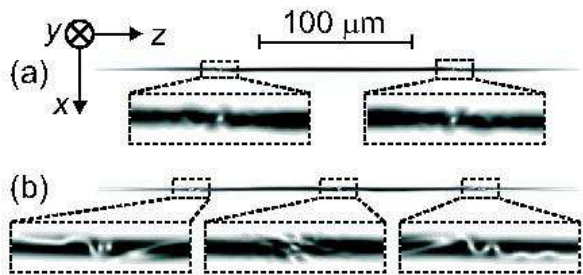}
\caption{(a) Atom density profiles in the $y=0$ plane (axes inset) of merged BECs ($t=\tau = 5$ ms) calculated, without fluctuations, for $\Delta = 4 \times 10^{4}z$ (a) and $\Delta=0.05\pi\sin(10{^4}\pi z) + \pi$ (b). Large dashed rectangles contain enlargements of density profiles within smaller rectangles. Horizontal bar shows scale.}
\label{f5}
\end{figure}

We can investigate further the effect of a spatially-varying phase on the merger by imposing a particular $z-$dependence on $\Delta$ at $t=0$. Firstly, we consider a linear phase gradient, $\Delta = 4 \times 10^{4}z$, which means that, initially, the two BECs are moving with a relative speed $v = 4 \times 10^{4}\hbar/m=0.1$ mm s$^{-1}$ in the $z-$direction. Since this speed is so low, the phase gradient has a negligible effect on the center-of-mass position of the approaching BECs when $\tau=5$ ms. However, the condition for resonant soliton production, $\Delta = \pi$, is only satisfied for $z=\pm 79$ $\mu$m. As a result, the merged BEC contains vortices at these positions only [Fig.~\ref{f5}(a)]. Consequently, the initial relative velocity $v=h/m \delta z$ can be determined by measuring the distance $\delta z$ between vortices in the merged cloud. This distance is $\sim 300$ times larger than the change in the atoms' center-of-mass position that the initial phase gradient produces over time $\tau$, thus magnifying the BECs' relative motion. Furthermore, the direction of vortex rotation indicates the direction of relative motion: positive $\partial \Delta / \partial z$ creates anticlockwise rotation in the $y=0$ plane, whilst negative $\partial \Delta / \partial z$ creates clockwise rotation in the $y=0$ plane. Non-linear phase gradients can also be studied using this technique. Figure~\ref{f5}(b) shows a merged cloud at $t = \tau = 5$ ms, for a sinusoidal phase imprint $\Delta= 0.05\pi\sin(10{^4}\pi z) + \pi$, like that created in a diffraction experiment~\cite{gunther}. Vortices are now observed at $z=0$ and $\pm100$ $\mu$m. 

In summary, we have shown that the decay of a dark soliton into vortices is the physical mechanism that enables the extraction of phase information in recent experiments on atom chip interferometry~\cite{kettnew}. The design of the interferometer can be tailored for different purposes. If all values of $\Delta$ are to be measured, the atom clouds should be at finite temperature and elongated to enhance phase fluctuations, and hence cause a smoother dependence of vortex production on $\Delta$. However, a very precise detection of $\Delta = \pi$ may be required, as, for example, in the ``motion detector'' shown in Fig.~\ref{f5}(a). In this case, either the BECs should be very cold, or approximately spherical, in order to suppress phase fluctuations. 

We thank W. Ketterle for helpful discussions. This work is funded by EPSRC.
%This precise measurement of BEC motion could be used to probe weak forces such as surface potentials or gravity. 

%This work is funded by EPSRC UK.
%\vspace{-0.2 cm}
% plain unsrt alpha abbrv
%\renewcommand\bibname{References}
%\nocite{*}
%\bibliographystyle{unsrt}
\bibliography{biblio}

\begin{thebibliography}{21}
\bibitem[{foo({\natexlab{a}})}]{foot1}
\bibinfo{note}{Strictly, we should keep our cutoff wavevector small and use a
  harmonic oscillator basis to ensure macroscopic thermal occupation of the BEC
  modes~\cite{davis}. Consequently, we do not attempt to ascribe a specific
  temperature value to our simulations.}
\bibitem[{foo({\natexlab{b}})}]{foot2}
\bibinfo{note}{As the vortices are approximately cylindrically symmetrical
  along the $y$-direction, their associated Fourier components are almost
  identical in the $x$- and $z$-directions.}
\end{thebibliography}

\end{document}